\documentclass[twoside,12pt]{article}
\usepackage{epsfig}

\def\Journal#1#2#3#4{{#1} {#2} (#4) #3 }

\def\NPA{{\em Nucl. Phys.} A}

\def\PPNP{{\em Prog. Part. Nucl. Phys.}}

\def\PLB{{\em Phys. Lett.} B}

\def\PRL{\em Phys. Rev. Lett.}

\def\PREP{\em Phys. Rep.}

\def\PRC{{\em Phys. Rev.} C}

\def\EPJA{{\em Eur. Phys. Jour.} A}

\newcommand{\be}{\begin{equation}}
\newcommand{\ee}{\end{equation}}
\newcommand{\bea}{\begin{eqnarray}}
\newcommand{\eea}{\end{eqnarray}}

\begin{document}

\title{ \vspace{1cm} The High-Density Symmetry Energy in Heavy Ion Collisions}
\author{H. H.\ Wolter,$^{1,4}$ V.\ Prassa,$^2$  G.\ Lalazissis,$^{2}$
T. Gaitanos,$^3$\\
G.\ Ferini,$^4$ M.\ Di Toro,$^{4,5}$ V.\ Greco,$^{4,5}$
\\
$^1$ Fakult\"at f\"ur Physik, Ludwig-Maximilians-Universit\"at M\"unchen, Germany\\
$^2$ Physics Dept., Aristotle University of Thessaloniki, Greece\\
$^3$ Inst. of Theoretical Physics, Justus-Liebig Univ, Giessen, Germany\\
$^4$ INFN, Laboratori Nazionali del Sud, Catania, Italy\\
$^5$ Dip. di Fisica e Astronomia, Universita degli Studii, Catania, Italy
}
\maketitle
\begin{abstract}
The nuclear symmetry energy as a function of density is rather poorly constrained theoretically and experimentally both below saturation density, but particularly at high density, where very few relaevant experimental data exist. We discuss observables which could yield information on this question, in particular, proton-neutron flow differences, and the production of pion and kaons in relativistic heavy ion collisions. For the meson production we investigate particularly ratios of the corresponding isospin partners $\pi^-/\pi^+$ and $K^0/K^+$, where we find that the kaons are an interesting probe to the symmetry energy. In this case we also discuss the influnece of various choices for the kaon potentials or in-medium effective masses.
\end{abstract}

\section{Introduction}
The symmetry energy $E_{sym}$ is that part of the nuclear equation of state (EOS), which depends on the isovector density $\rho_3=\rho_p-\rho_n$,
$\epsilon(\rho,\rho_3)=\epsilon(\rho)+\frac{E_{sym}}{A}\rho_3^2/\rho+\cdots$ \cite{baran_PRep}. It has been found that the quadratic approximation in $\rho_3$ is a very good assumption at a wide range of density (but may break down at very low density due to the importance of cluster correlations). While the symmetry is fairly well constrained around saturation density from the Bethe-Weizs\"acker mass formula, its behavior  both at lower and at higher densities is of great recent interest both theoretically  and experimentally \cite{fuchs_wci, BALi_Schrod}. At low densities it is important for the structure of exotic nuclei as well as in the neutron star crust and in supernova explosions. For high densities it is crucial for the structure of neutron stars, in particluar as to the question, whether strange or deconfined hadronic matter exists in the center of neutron stars \cite{klaehn}.

Heavy ion collisions provide a unique opportunity to explore the density dependence of the symmetry energy in the laboratory, because one is able to choose the asymmetry via the collision partners, and the density range via the collision energy and centrality. On the other hand, one does not measure the equation of state, and thus also not the symmetry energy, directly in a heavy ion collision, which is a violent non-equilibrium process. Rather on has to perform transport calculations of the collision process, into which the EOS (and the in-medium cross section, which are also isospin-dependent) enter, and compare observables which are sensitive to the symmetry energy with experiment. The search for such sensitive observables is one of the great challenges in this field.

A broader view of the investigation of the symmetry energy in heavy ion collisions is given in the contribution of M. Di Toro at this conference \cite{ditoro_eric}. There is also an extensive recent review on the symmetry energy in heavy ion collisions by B.A. Li, et al. \cite{BALi_PRep}. Here we concentrate on the high density symmetry energy, which is investigated in relativistic heavy ion collisions. The observables discussed here, are, of course, the same which are discussed to investigate the EOS generally, i.e. the symmetric part of the energy density: flow observables, pre-equilibrium emission of nucleons and light clusters, and the production of mesons. However, here we are looking for isospin differences, i.e. proton/neutron flow differences, or differences in the production of isospin partners, e.g. $\pi^+$ vs.  $\pi^-$ or $K^0$ vs. $K^+$. It is clear, that these differences will be much smaller, than effects of the symmetric EOS, and thus the signatures of the symmetry energy are difficult to grasp. This may also be the reason, why only very few experimental data are available et present, which look directly for isospin differences. However, this situation should improve in the future, when more intense radiactive beams become available at facilities such as FAIR, RIKEN, FRIB, etc.

\begin{figure}[t]
\begin{center}
 \parbox{8cm}{\includegraphics[width=8.5cm]{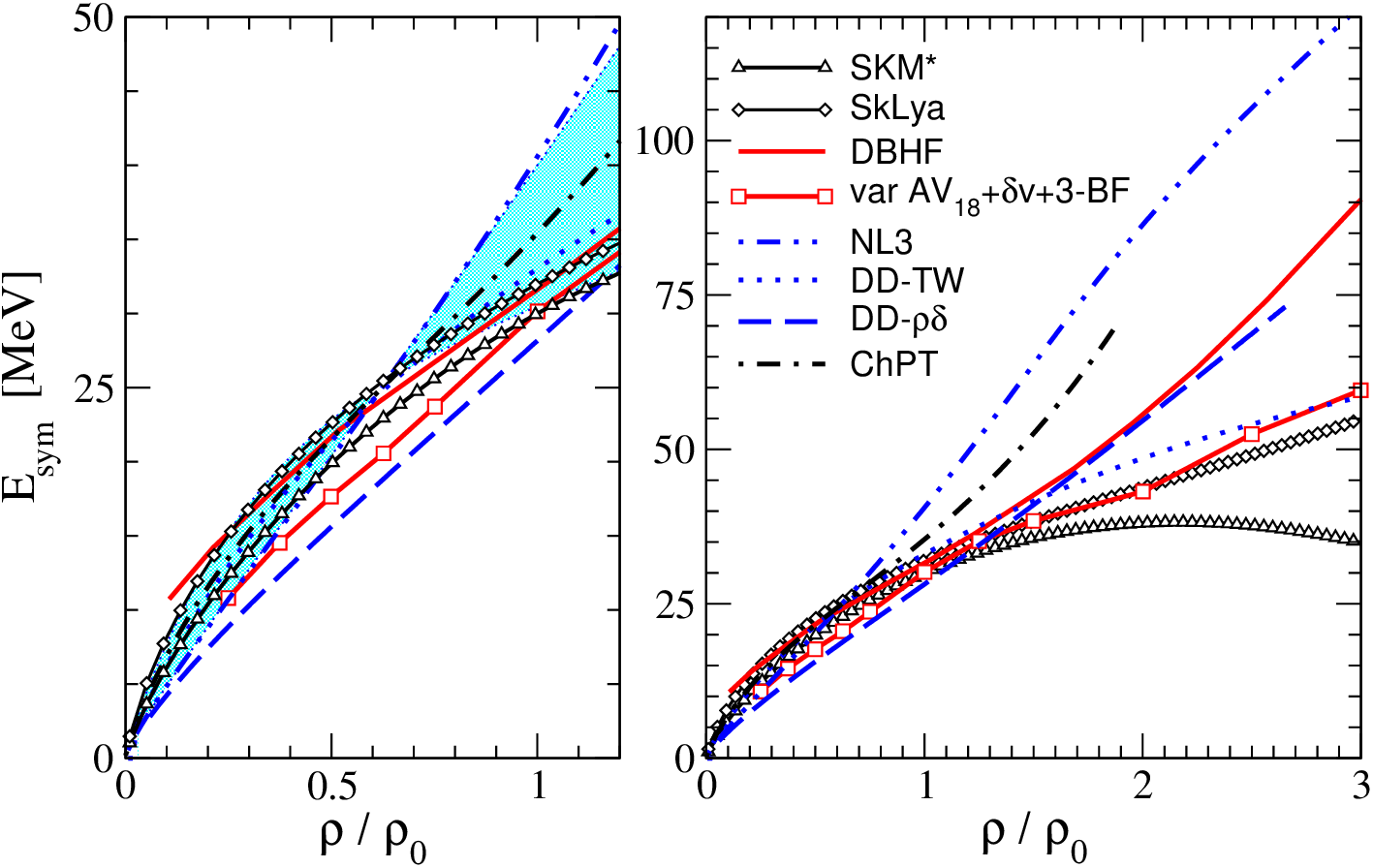}}
 \hspace*{0.3cm}
 \parbox{9cm}{\includegraphics[width=8.5cm]{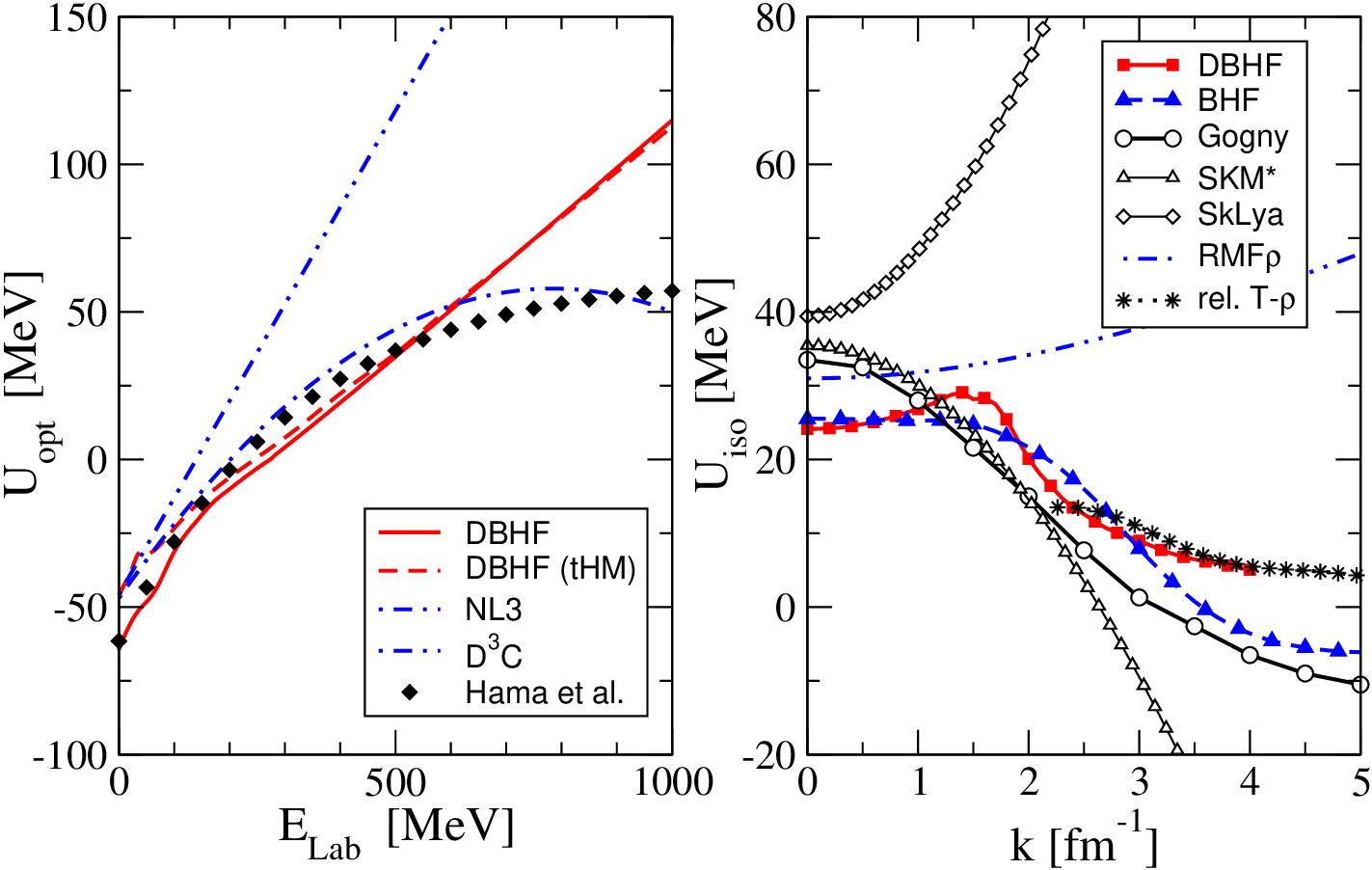}}
 \caption{
 (a-d from left to right): (a,b) Symmetry energy as a function of 
normalized density in various theretical nuclear matter models (panel(a) detail for low density). The theoretical models are labeled in panel (b), and and given explicitely in ref.\cite{fuchs_wci}.
\newline
 (c) Real part of optical potential for symmetric nuclear matter as a function of energy, and (d) isospin part of optical potential (Lane potential) for asymmetric nuclear matter as a function of momentum, in some of the theoretical models identified in the panels, see ref.\cite{fuchs_wci}.}
\label{symm_energ}
\end{center}
\end{figure}

\section{Models of the Symmetry Energy and Transport}

Theoretical models for nuclear matter have been formulated non-relativistically or relativistically in phenomenological models (Skyrme-type, resp. Relativistic Mean Field (RMF)) or in microscopic models (Brueckner HF or Dirac-Brueckner HF (DBHF)). A connection between the phenomenological and the mircroscopic approaches can be seen in the density functional approach: the density functional of the phenomenological models is guided (or taken directly) from the density dependence of the mean field of the microscopic models. These observations are true for the symmetric EOS as well as for the symmetry energy. However, the symmetric EOS is much more constrained from properties of nuclear matter and finite nuclei, than the symmetry energy. Consequently there is a large theoretical uncertainity about the density behavior of the symmetry energy.

This is demonstrated in Fig.1: In panels (a) and (b) the symmetry energy is given in various phenomenological and microscopic models (detailed references to these models are found in ref.\cite{fuchs_wci}, from which these figures are taken). One may distinguish between symmetry energies which rise more or less strongly around saturation density (asy-stiff, resp. asy-soft). It is seen that most of the realistic models cross not at $\rho_0$, but rather around $0.6 \rho_0$. This points to the fact, that in fitting asymmetric nuclei the asymmetry of the surface has a large influence. For densities below saturation the realistic models of the symmetry energy thus do not deviate very strongy; however, this still has important consequences, as demonstrated in ref.\cite{ditoro_eric}. It is clear, however, that dramatic differences in the symmetry energy exist for higher densities. One may also notice, that the behaviour of the symmetry energy for low and high density is not necessarily connected. E.g. the symmetry energy in DBHF calculations is rather soft for low densities but behaves stiff at higher densities \cite{fuchs_wci}.

In Fig.1, panels (c) and (d), the energy, resp. momentum dependence of the real part of the optical potential is shown for symmetric nuclear matter (c) and the Lane potential for asymmetric nuclear matter from various calculations \cite{fuchs_wci}. In (c) the calculations are compared to the empirical potential (Hama et al.), and it seen that the energy dependence is too strong in RMF (NL3) or DBHF models, but can be reproduced in a recently developed RMF model with density and derivative dependent coupling (D$^3$C, \cite{typ_D3C}). Again the Lane potential shows much larger deviations already for not so high momenta. Experiments indicate that the Lane potential is decreasing with energy, but the accuracy is not enough to essentially constrain the momentum dependence of the symmetry energy. The momentum dependence of the symmetry energy can also be dicussed in terms of effective masses, here specifically of the difference of proton and neutron effective masses \cite{fuchs_wci,ditoro_eric}.

\begin{figure}[t]
\begin{center}
\includegraphics[width=8.5cm]{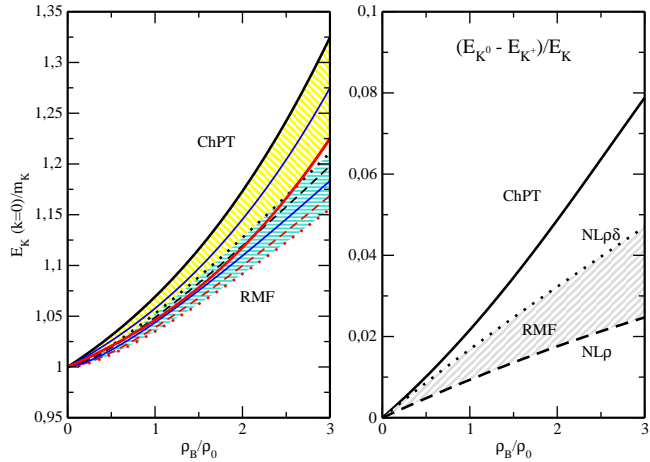}
\caption{(left) Density dependence in asymmetric nuclear matter ($\alpha=0.2)$ of the in-medium kaon mass in units of the free kaon mass for the ChPT model \cite{kaon_chpt} (yellow) and the OBE/RMF model \cite{kaon_obe} (blue). The upper/lower curves are for $K^0 / K^+$, the middle one for symmetric nuclear matter. For RMF the kaon mass also depends on the EOS ($NL\rho$ dashed, $NL\rho\delta$ dotted). (left) Relative isospin splitting for the same models.}
\label{kaon_mass}
\end{center}
\end{figure}

Heavy ion collisions are simulated in this work, using a relativistic transport code with finite size test particles, for details see ref.\cite{RLV}. We propagate protons and neutron separately, but also particles produced in inelastic NN collisions, such as $\Delta$ resonances, pions and kaons. The self energies for the nucleons are specified in the RMF model, which includes non-linearity in the $\sigma$-field, and for the isovector sector either $\rho$ mesons, or $\rho$ and $\delta$ mesons (models $NL\rho$ and $NL\rho\delta$, respectively) \cite{gait_Lorentz}. An issue has been the specification of the mean field for the kaons \cite{fuchs_kaon}. It has been modelled in chiral perturbations theory (ChPT, Nelson and Kaplan, \cite{kaon_chpt}) or in the one-boson exchange model (OBE, \cite{kaon_obe}), and we test both options here, as discussed below (for details see ref.\cite{kaon_NPA}). Here we discuss in particular the isospin partners of the strange kaons ($K^+, K^0)$, since the anti-strange kaons have a much larger final state interaction. The effective kaon mass for the two models is shown in Fig, 2; on the left side for the different kaon states, and on the right hand side for the relative change, which is seen to be much larger in the Nelson-Kaplan model.

\section{Observables for the High-Density Symmetry Energy}

In this section we want to discuss observables which could constrain the high density symmetry energy. As already noted in the Introduction, there are very few relevant experimental data at present, and thus we want to explore how to further investigate symmetry effects at relativistic energy. Here we discuss collective flow and the properties of produced pions and kaons. Another possible signal are ratios of protons to neutron or of light clusters, which are emitted in the early stage of the collision (pre-equilibrium particles). These are in a sense complementary to the bulk flow of protons and neutrons. Pre-equilibrium particles are discussed also at lower energy in ref.\cite{ditoro_eric}.

\begin{figure}[t]
\begin{center}
 \parbox{8cm}{\includegraphics[width=8.5cm]{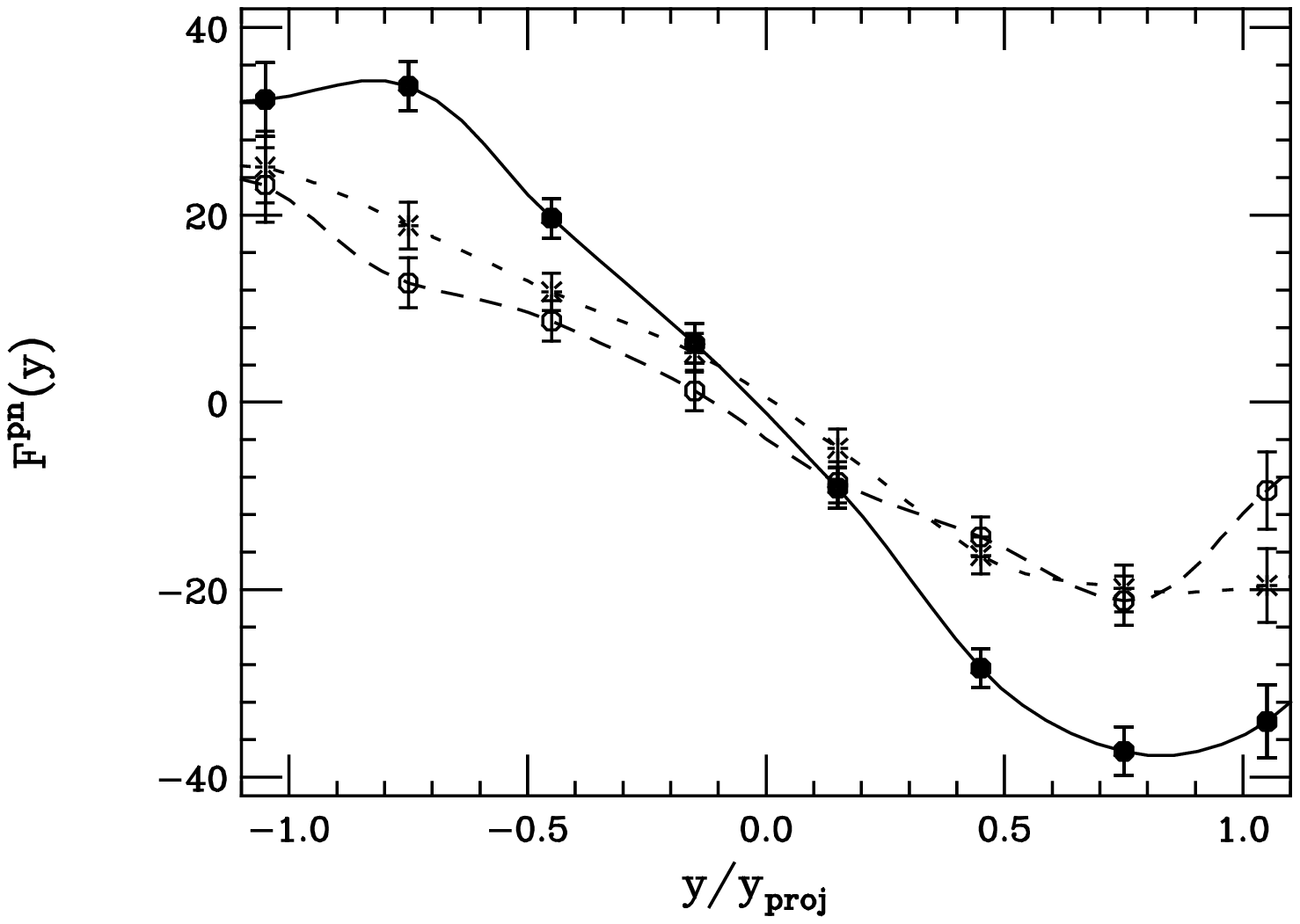}}
 \hspace*{0.3cm}
 \parbox{9cm}{\includegraphics[width=8.0cm]{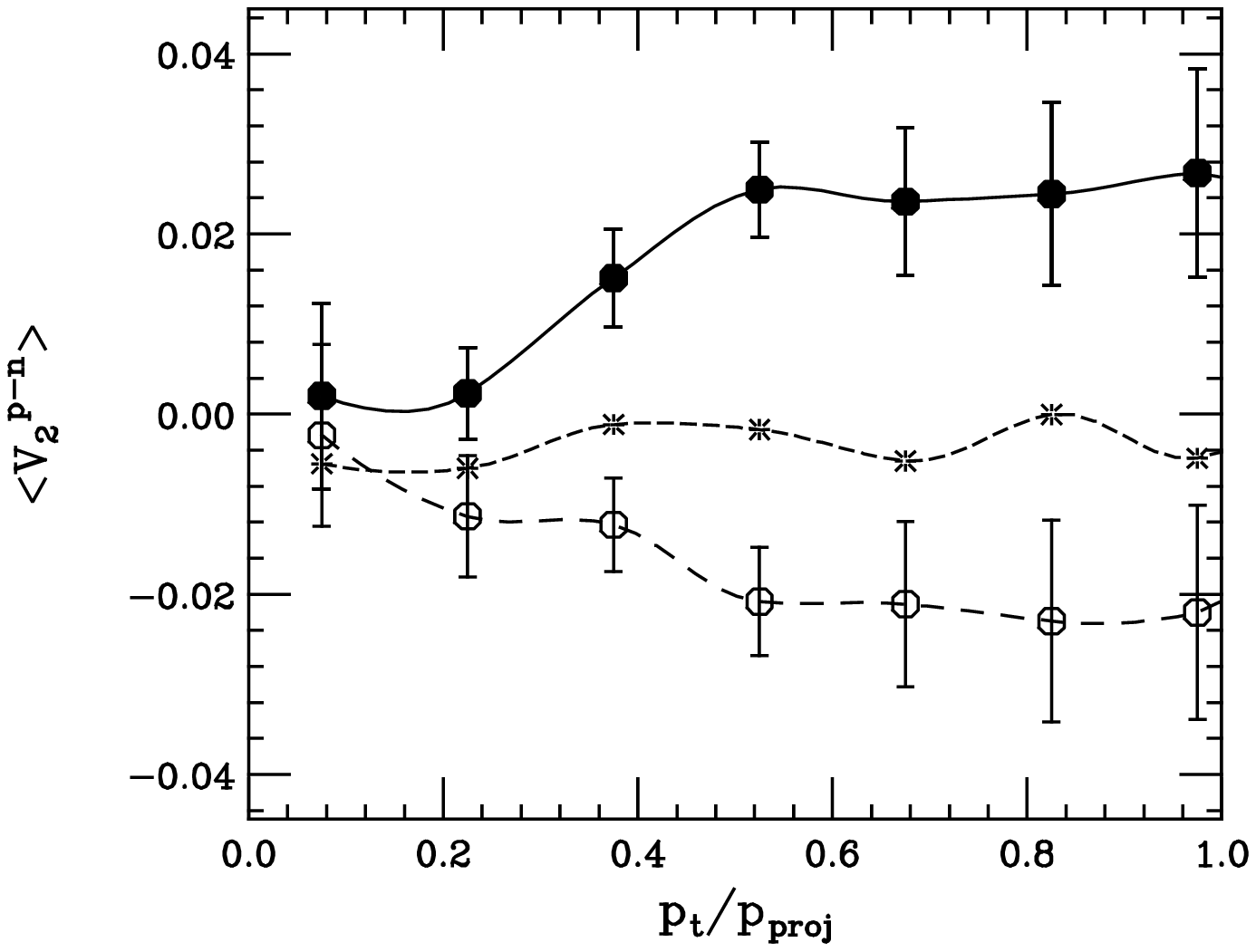}}
 \caption{
 Proton-neutron differential directed flow ($p^x$) (left) and elliptic flow ($v_2$) (right) for the collision $^{132}Sn+^{132}Sn$ at 1.5 AGeV incident energy as a function of the normalized rapidity \cite{isoflow}. The different curves are for the model $NL\rho\delta$ (solid), $NL\rho$ (dashed) and without a symmetry potential ($NL$, dotted). }
\label{pn_diff_flow}
\end{center}
\end{figure}

\subsection{Proton-Neutron Differential Flow}

The mean field directly influences the momentum distribution of the particles in the final state of the collision, and thus the so-called flow observables. The yield is a function of rapidity, transverse momentum and azimuthal angle, and has been expanded into a Fourier series in the azimuthal angle. The lowest coefficients are the directed ($v_1$) and elliptic ($v_2$) flow.  These investigations have been crucial to determine the EOS of symmetric nuclear matter. The symmetry energy gives rise to different forces for neutrons and protons (repulsive for neutron in a neutron-rich medium, and attractive for protons). It is thus the difference of proton and neutron flow, which should be sensitive to the symmetry energy, or also the proton-neutron differential flow, defined as $F^i_{p-n} = 1/N \sum_n{p^i_n \tau_n}, \tau_n=\pm 1$ for neutrons/protons \cite{isoflow}.

Results for the directed and elliptic proton-neutron differential flow are shown in Fig.3 for the collision $^{132}Sn+^{132}Sn$ at 1.5 AGeV incident energy. Here the isospin-dependent RMF models $NL\rho$ and $NL\rho\delta$ have been used for demonstration, where the latter model has the stiffer symmetry energy. It is clearly seen that there is an appreciable signal in these flow observables, which is still present when the more realistic case as $^{132}Sn+^{124}Sn$ is considered \cite{isoflow}. In fact, it can be easily demonstrated, that the differential flow is a direct consequence of a cancellation of the large isovector vector ($\rho$) and isovector scalar ($\delta$) fields, analogously to the cancellation of the $\sigma$- and $\omega$-fields in the isoscalar sector \cite{isoflow}. A direct measurement of isospin differences of flow observables would highly desirable.

\begin{figure}[t]
\begin{center}
 \parbox{7cm}{\includegraphics[width=7.2cm]{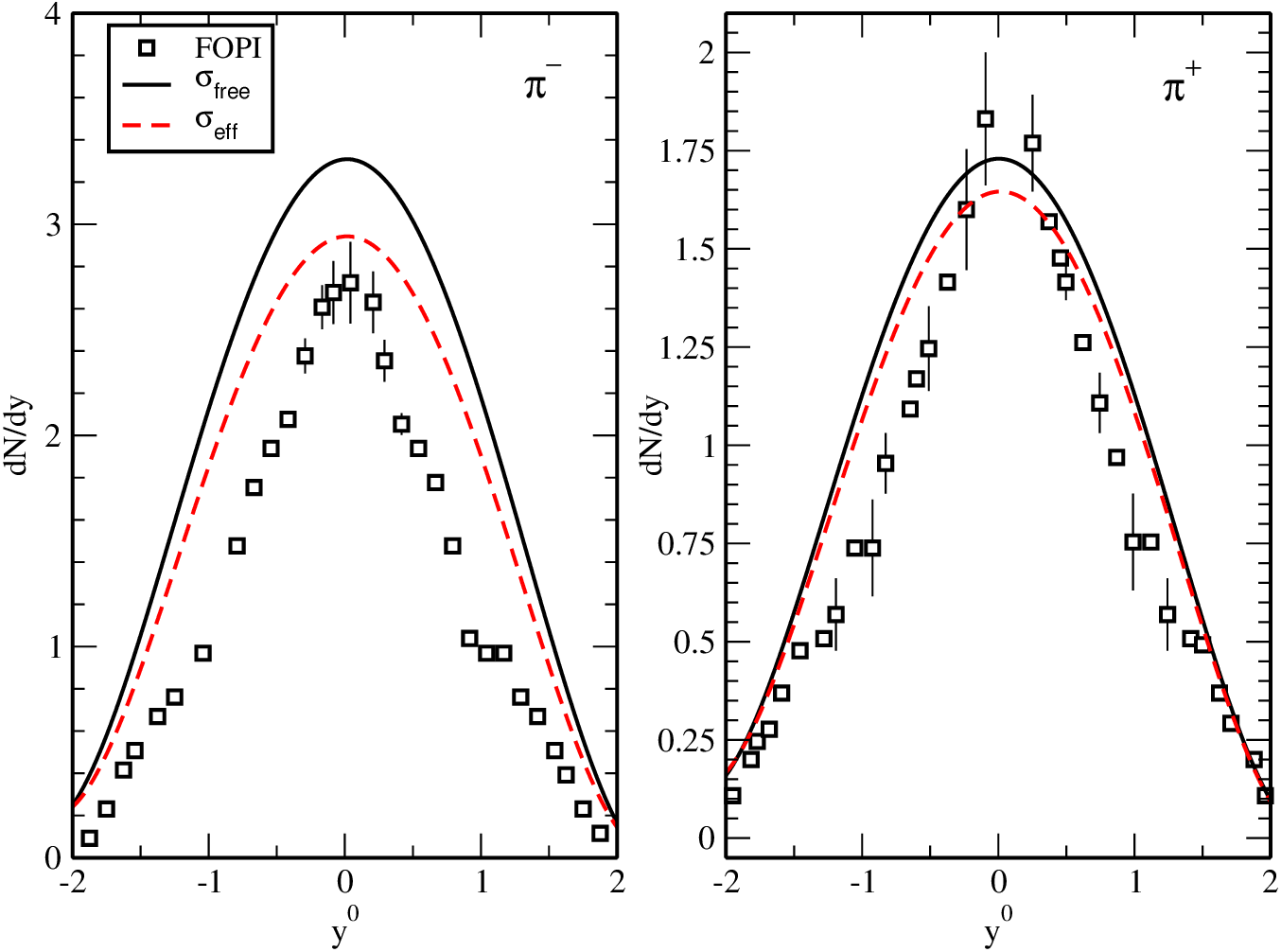}}
 \hspace*{0.3cm}
 \parbox{9cm}{\includegraphics[width=9cm]{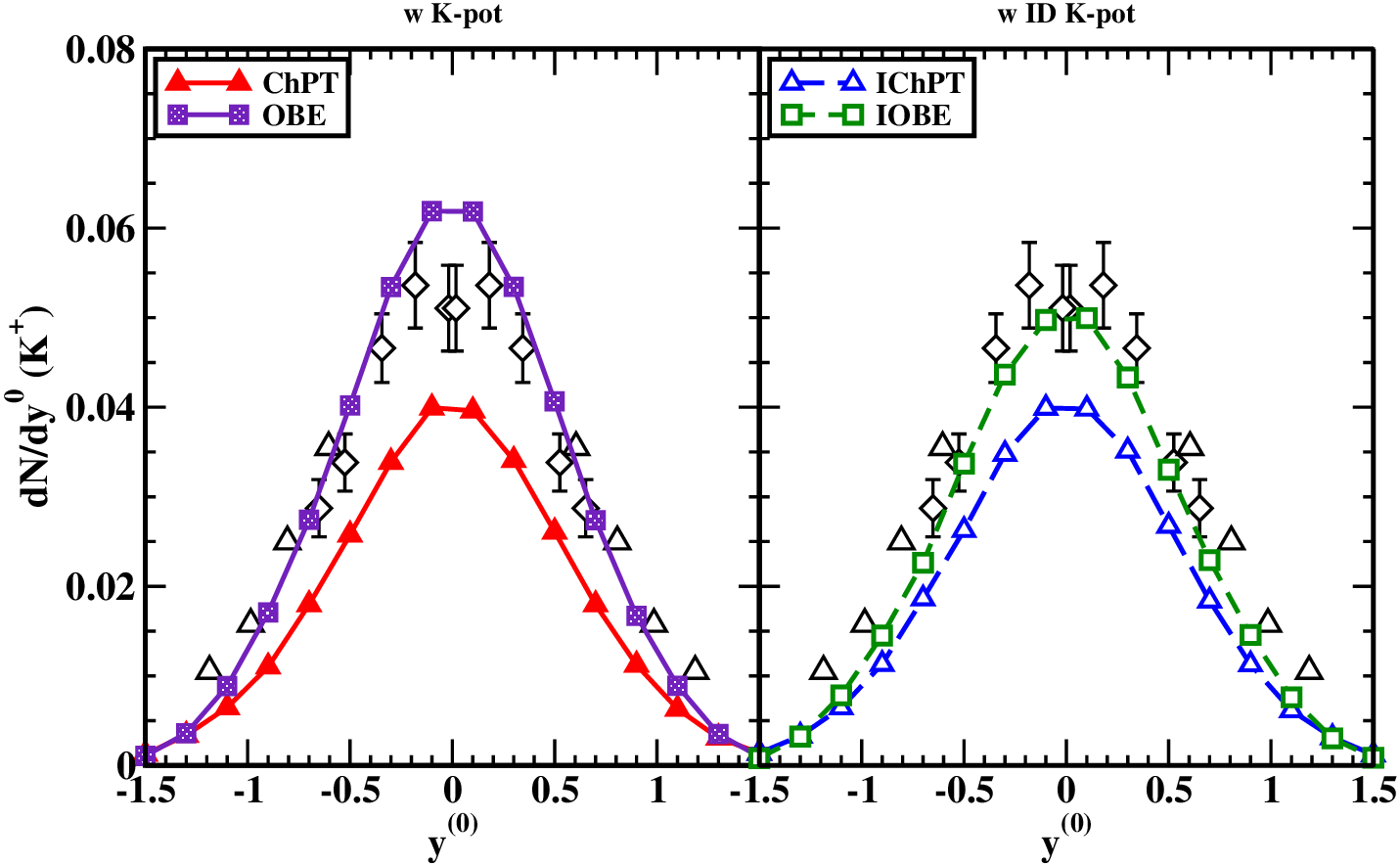}}
 \caption{
(left) Rapidity distributions of $\pi^-$ and $\pi^+$ in semicentral $Au+Au$ collisions at 800 AMeV in comparison with data from FOPI \cite{reisd_pi}, using free (solid) or medium modified (dashed) inelastic $NN->N\Delta$ cross sections.
\newline
(right) Rapidity distributions of $K^+$ mesons in semicentral $Ni+Ni$ collisions at 1.93 AGeV, in comparison with data from FOPI \cite{fopi_K} and KaoS \cite{kaos_K}. Theoretical calculations are given for the isospin-independent part of the kaon potential in the ChPT and OBE models (left), and including the isospin dependent part (right) \cite{prassa_Klet}.
 }
\label{dndy_piK}
\end{center}
\end{figure}

\subsection{Meson Production}

Meson are produced as secondary particles in a heavy ion collisions and can serve as a probe of the dense matter created transiently in the collision. At relativistic energies pions are generated copiously via the production and subsequent decay of the $\Delta$ resonance. However, pions continue to interact strongly with nucleons also in the later phases of the collisions and their properties depend on the complete evolution of the collision. Strange mesons, kaons, on the other hand are produced sub- and near-threshold in this energy range, and the main mechanism is via the secondary production from $\Delta$s and pions. Therefore kaons are particularly a probe of the dense phase of the collision. Strange kaons (in contrast to anti-strange kaons) have the additional advantage to interact weakly with nucleons in the final state, and thus constitute a particularly promising probe of the matter in heavy ion collisions. In fact, the ratio of positive kaons produced in collisions of heavy ($Au+Au$) relative to light ($C+C$) systems has been one of the most sensitive probes for the EOS of symmetric nuclear matter \cite{fuchs_PRL}.

It is thus an interesting question whether the relative production of the isospin partners of pions and kaons is sensitive to the density dependence of the symmetry energy. In fact, the ratio of the yields $\pi^+/\pi^-$ depends on the isospin asymmetry of the matter, where they are produced, and consequently also the ratio $K^0/K^+$. For kaons, in addition, the mean field potential or in-medium effective mass is important for the threshold condition in the near-threshold production \cite{ferini_kaon}, as was discussed above in connection with Fig.2. In fact, in this way one may obtain additional information on the still uncertain kaon potential.

\begin{figure}[t]
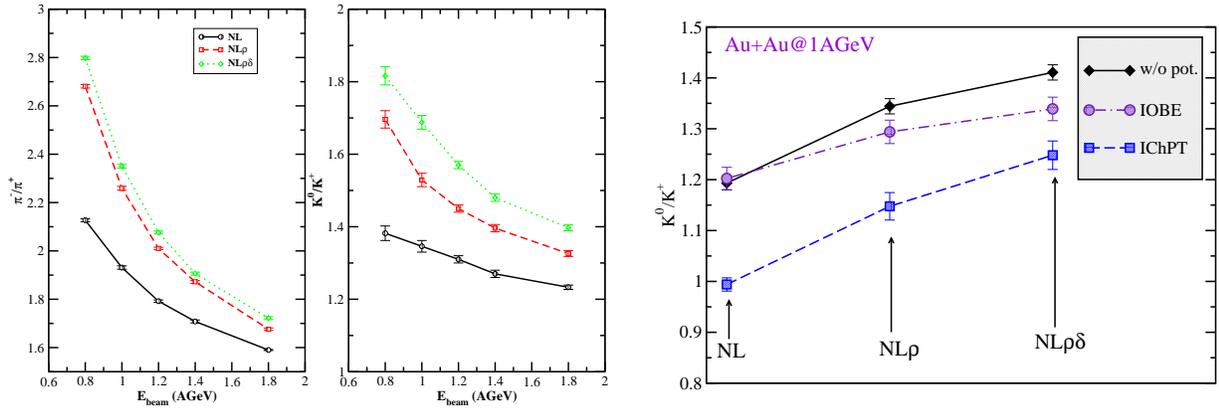

\begin{center}
 \parbox{8cm}{\includegraphics[width=8cm]{wolter_fig5left.eps}}
 \hspace*{0.3cm}
 \parbox{8cm}{\includegraphics[width=7.5cm]{wolter_fig5right.eps}}
 \caption{
(left) Ratio of $\pi^-/\pi^+$ (left) and $K^0/K^+$ (right) in $Au+Au$ collisions as a function of incident energy. Shown are calculations with   various choices of the symmetry energy: $NL$ (solid), $NL\rho$ (dashed), and  $NL\rho\delta$ (dotted). For the kaons only the isospin-independent part of the ChPT model is used.
\newline
(right) Kaon yield ratio in central $Au+Au$ colisions at 1 AGeV for the varios choices of the symmetry energy (as in Fig.5) but for different assumptions of the kaon potential: w/o kaon potential (solid), with the iisospin-dependent kaon potential in the OBE (dash-dot) and the ChPT (dashed) models.
 }
\label{piK_ratio}
\end{center}
\end{figure}

In Fig.4 we show rapidity distributions of pions (left) and kaons (right) in comparison with data \cite{prassa_Klet}. For pions we see that the yield is generally underpredicted, in line with the general observation that the total yield of pions is generally underpredicted in calculations \cite{reisd_pi}. However, we also see, that the agreement is reasonable at mid-rapidity, which corresponds to the production at the highest densities and thus to the condition where kaons are most likely produced. Thus one can expect the description of kaon production to be reasonable. In fact, in the right panels, the rapidity distributions of kaons are generally well described. One does see, however, that the kaon potential has a substantial influence on the kaon yields. There are effects both of the model for the kaon potential (ChPT or OBE) and also of the isospin-dependent parts of these potentials \cite{prassa_Klet}. It was seen in Fig.2 that the isospin dependent part is larger in the ChPT model, and thus it affects the yields more strongly.

It is therefore attractive to consider ratios of pion ($\pi^+/\pi^-$ ) or kaon ($K^0/K^+$) yields in order to cancel out some of the unknown behavior and have a clearer signature of the symmetry energy \cite{kaon_PRL}. This is shown in Fig.5 as a function of incident energy for $Au+Au$ collisions for different choices of the symmetry energy (and without the isospin-dependent part of the kaon potential). It is seen that the effect of different choices for the symmetry energy is rather small for the pion ratio. This points to the fact that pions are produced during the whole evolution of the collision and not necessarily at high density where the symmetry energy differences are large. On the other hand there is a substantial effect for the kaon ratio, because of their sensitivity to the high-density phase.

There exist data of the FOPI collaboration of the double ratio of $K^0/K^+$ for the systems $Ru+Ru/Zr+Zr$ in comparison with calulations \cite{lopez_fopi}. These show a rather small effect of the symmetry energy for calculations in collisions of finite nuclei. This was in contrast to calculations in infinite nuclear matter, where a large sensitivity was predicted \cite{ferini_kaon}. The reason was seen to be that the asymmetry of the initial system changes rather fast in the collision and thus washes out the effects of the symmetry energy. But it should perhaps be more conclusive to measure the ratios for one asymmetric system absolutely.

It is of interest to see whether the details of the kaon potential influence the sensitivity to the symmetry energy. This is tested in Fig.6, where the kaon ratios $K^0/K^+$ are shown for the different choices of the symmetry energy, and for different specification of the kaon potential. The ratios are not sensitive to a kaon potential per se (the curves with isospin independent potential practically coincide with those without kaon potential), but there is an influence for the isospin dependent part of the kaon potential, particularly for the stronger effect in the ChPT model. Thus, one also may obtain information on this rather uncertain quantity.

\section{Concluding Remarks}

In this contribution we investigated observables in relativistic heavy ion collisions which could yield information on the high density symmetry energy. By nature the signatures are smaller for the symmetry energy (which represents only a small part of the total mean field) than the corresponding ones for the symmetric part of the EOS, and they are thus more difficult to measure. However, proton-neutron differences of flow are of interest, if neutrons can be measured simultaneously. Also ratios of light clusters can yield significant signatures, as is shown for the effect of the momentum dependence of the symmetry energy in ref.\cite{ditoro_eric}. On the other hand, pion and particularly kaon ratios present a promising signature, in particularly if they can be measured absolutely for an asymmetric system. On the other hand, important additional information comes from observations of neutron star properties on the high density symmetry energy, as has recently been discussed in ref.\cite{klaehn}.

{\bf Acknowledgements}:
This work was supported in part by the Greek State Scholarshop Foundation (IKY) and by the German DFG Cluster of Excellence {\em Origin and Structure of the Universe} ({\em www.universe-cluster.de}).

\end{document}